# Holding Dissapearance in RTD-based Quantizers

*Juan Núñez, José M. Quintana and María J. Avedillo*
Instituto de Microelectrónica de Sevilla, Centro Nacional de Microelectrónica,
Edificio CICA, Avda. Reina Mercedes s/n, 41012-Sevilla, SPAIN
FAX: +34-955056686, E-mail: {jnunez, josem, avedillo}@imse.cnm.es

**ABSTRACT**

Multiple-valued Logic (MVL) circuits are one of the most attractive applications of the Monostable-to-Multistable transition Logic (MML), and they are on the basis of advanced circuits for communications. The operation of such quantizer has two steps: sampling and holding. Once the quantizer samples the signal, it must maintain the sampled value even if the input changes. However, holding property is not inherent to MML circuit topologies. This paper analyses the case of an MML ternary inverter used as a quantizer, and determines the relations that circuit representative parameters must verify to avoid this malfunction.

## 1. INTRODUCTION

Resonant tunneling devices are today considered the most mature type of quantum-effect devices, already operating at room temperature, and being promising candidates for future nanoscale integration. Resonant tunneling diodes (RTDs) use to be implemented in III-V materials, and they exhibit a negative differential resistance (NDR) region in their current-voltage characteristics which can be exploited to significantly increase the functionality implemented by a single gate in comparison to MOS and bipolar technologies [1].

Figure 1a shows the *I-V* characteristic of a RTD enhancing key parameters for circuit design: peak current and voltage, $I_p$ and $V_p$, and valley current and voltage, $I_v$ and $V_v$. Three regions are defined according to Fig. 1a: two regions of positive (I and III) and one of negative (II) differential resistance. Circuit applications of RTDs are mainly based on the MOnostable-BIstable Logic Element (MOBILE) [2-4]. The basic MOBILE is a rising edge triggered current controlled gate which consists of two RTDs (the load and driver RTDs) connected in series and driven by a switching bias voltage ($V_{bias}$). When connected in series, RTDs provide multiple-peak structures in their *I-V* characteristics, which make it attractive for multiple-valued logic (MVL). MVL circuit applications are based on the Monostable-to-Multistable transition Logic (MML) [5], an extension of the binary MOBILE. Logic operation is based on the sequential switching of the RTDs connected in series, which is produced when the bias voltage rises to an appropriate value. Logic functionality is achieved by embedding an input stage (compound-semiconductor transistors, HEMT or HBT) which modifies, according the applied input signal, the peak current of some of the RTDs.

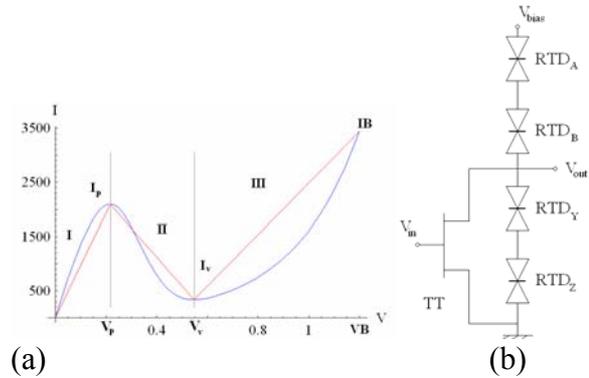

**Figure. 1**. (a) *I-V* characteristic for a LOCOM RTD (blue) and its linear approach (red) and (b) typical structure of a MML ternary inverter.

Multiple-valued quantizers are considered one of the most attractive applications of MML circuits, and due to their high-speed performance, they are expected to be exploited in future commercial applications of communication systems [6], [7]. The sample & hold quantizer has the particularity that samples the input whenever it receives a trigger event (the raising of the bias voltage) and holds the output at the acquired input value until the next triggering event occurs. This holding property, we refer as *holding*, is crucial to guarantee a correct operation of this kind of structure.

In this paper, we consider a sample & hold quantizer which has been implemented by using a ternary-valued inverter and we study the holding problem associated with it. The analysis performed shows that this is not an inherent property to the circuit topology. The difficulty of analytically studying the circuit has been overcome by resorting to simplified (piecewise linear) descriptions for the RTD driving point characteristics. Finally, relations between RTD and transistor parameters that ensure a correct behavior have been obtained.

Models for RTDs and transistors from LOCOM [8] have been used. For this RTD, $V_p$ is 0.21V, the peak current density 21 KA/cm$^2$, and the peak current ratio is about 6.25 at room temperature. The transistor is a depletion HFET with threshold voltage -0.2V. In order to simplify the algebraic analysis, we have considered a linear approach of the RTD (Fig. 1a, in red) which has been defined through the currents and the voltages of the peak and the valley of the LOCOM model (in blue in Fig. 1a). This piecewise definition of the RTDs has been used to derive a linear multiple-peak I-V characteristic of the two series RTDs, which will provide an easiest way of analyzing the evaluation problem.

## 2. OPERATION PRINCIPLE

Figure 1b depicts a typical structure of a ternary inverter based on MML. The *driver* consists of a HFET (TT) and two series RTDs (RTD$_Y$ and RTD$_Z$). The HFET provides the logic functionality as its input modulates the drain to source current of the transistor, and consequently the total current through the driver. For this structure, three feasible values of the $V_{in}$ are considered, the high ($V_{in}^H$), medium ($V_{in}^M$) and low ($V_{in}^L$) voltages. The *load* is made up of the series connection of RTD$_A$ and RTD$_B$.

The operation of the MML inverter is determined by the relation between the peak currents of the driver and the load. If the input is at a high level, $V_{in}^H$, (ternary logic level "2"), the output must be at logic level "0". When $V_{in}=V_{in}^M$ (logic level "1"), the output must also be "1".

Finally, for the lowest level of the input ($V_{in}^L$), the output has to reach its highest value (level "2"). RTDs are supposed to have equal current densities so that peak currents are proportional to RTD areas.

Sizing the devices has a critical effect on the holding property of an MML inverter. Figures 2a and 2b depict HSPICE simulation results for an inverter with $f_Z$=0.6, $f_Y$=0.7, $f_B$=1.1 and $f_A$=1.2. The transistor form factor (*FF*=W/L) in Fig. 2a is *FF*=6, and the inverter does not present any holding problem because the output voltage keeps its logical level despite a variation of the input voltage. However, if *FF*=10, a malfunction appears when the input rises to the highest value (Fig. 2b) as the waveform marked with "2" indicates. When $V_{in}$ increases its value from $V_{in}^M$ to $V_{in}^H$, the output voltage does not hold its value and falls down to the lowest logical level.

## 3. THE HOLDING DISSAPEARANCE

The static behavior of the ternary inverter obeys the following expression:

$$g_L[V_{bias}^H - V_{out}] = g_D[V_{out}] + I_T[V_{in}, V_{out}] \qquad (1)$$

where $g[v]$ and $I_T[V_{GS}, V_{DS}]$ represent the mathematical description of two series connected linear RTDs ($g_D[v]$ for the driver and $g_L[v]$ for the load) and the transistor, respectively.
The holding disappearance problem comes from the disappearance of one (or more) of the stable states in the DC solution representation when $V_{bias}=V_{bias}^H$. A good analysis to what happens to this bad behavior can be done through the set of solutions to Eq. (1) in the $V_{in}$-$V_{out}$ plane. This plot depicts what happens when the input voltage changes its value while the bias voltage is set to $V_{bias}^H$, thus it is very useful to check the existence of solutions for each value of the input voltage. This representation provides a way of determining the critical condition from which the disappearance of solutions occurs, as will be studied in Section 4.

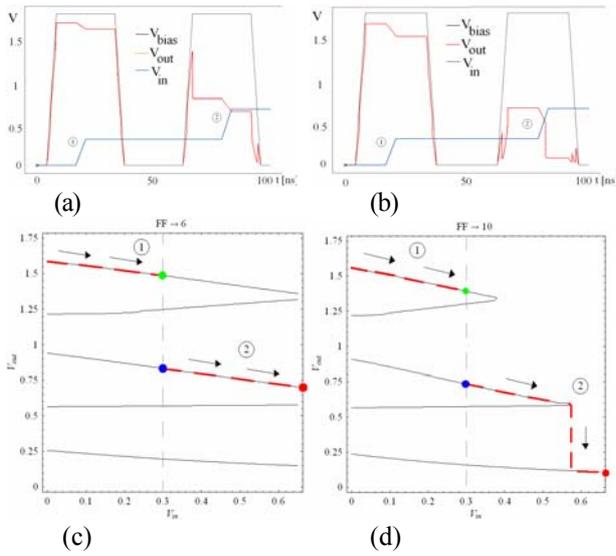

**Figure 2**. MML ternary inverter with $f_Z$=0.6, $f_Y$=0.7, $f_B$=1.1 and $f_A$=1.2. Waveforms for (a) *FF*=6 (does not exhibit holding problems) and (b) *FF*=10 (holding disappearance when the input node changes from $V_{in}^M$ to $V_{in}^H$). $V_{in}$-$V_{out}$ plots corresponding to (c) *FF*=6 and (d) *FF*=10.

Figures 2c and 2d depicts the plots corresponding to Eq. (1) for the HSPICE simulations of Figures 2a and 2b respectively. In the first one, a correct behavior is observed since the output level keeps its value despite of the variation of the input level while $V_{bias}=V_{bias}^H$. The red dotted line marked with "1" shows the evolution of the output voltage corresponding to a rise of the input from the lowest to the medium level. The vertical line is the input medium voltage level and the green point in Fig. 2c is the final output value. The second one (marked with"2") depicts such voltage when the input varies from the medium (blue point) to the highest level (red point). On the other hand, when the structure is not properly sized, holding problems could be found as Figure 3d points out. Paying attention to the red dotted line marked with "2" (variation of the input from $V_{in}=V_{in}^M$ to $V_{in}=V_{in}^H$), we can check that when the input voltage is about $V_{in}$=0.56*V*, the output falls down to the lowest level. The output does not maintain its value and the holding property is not verified.

To determine a criterion for a generic structure exhibits holding problems or not, we start from a qualitative analysis of the current-voltage characteristic and the $V_{in}$-$V_{out}$ representation, by which a relation between circuit parameters (area factor of the RTDs and form factor of the transistor) will be derived.

Figure 3a depicts the $V_{in}$-$V_{out}$ for $V_{in}=V_{in}^H$, marking with a red point the critical situation in which both solutions belonging to the highest value of the output collapse. This malfunction appears when the load is biased about its first peak voltage (RTD$_B$ is in its peak and commutes, forcing a change in the output node), thus, $V_{out}$ is approximately $V_{bias}^H - (V_{p1})_L$. Figure 3b plots the current-voltage characteristic of a MML inverter which has been sized properly to avoid this malfunction, where the red points and the grey area are used to check the existence of solutions of $V_{out}$ at high levels. The piecewise current-voltage characteristic of two series connected RTDs has been obtained from the linear approach of each individual RTD (see Fig. 1a). It is possible to calculate the peak and valley voltages and currents by simple geometrical considerations. This *I-V* characteristic does not include some regions that would appear in the complete representation corresponding to two linear series-connected RTDs, but they do not modify the normal operation of the inverter.

A malfunction is found around $V_{out} \cong (V_{p1})_D$ when $V_{in}=V_{in}^L$. It happens when RTD$_Z$ is in its peak voltage, increasing the value of the output node, which was set up to the highest level after the bias voltage reached $V_{bias}=V_{bias}^H$. If $V_{in}=V_{in}^M$, two holding problems must be analyzed. Figures 3e and 3f depict the first problem, which occurs around the maximum value of the output, that is, $V_{bias}^H - (V_{p1})_L$. Figures 3g and 3h plot the second one (near $V_{out} \cong (V_{p1})_D$).

## 4. CRITICAL DEPENDENCIES AND RESULTS

### 4.1 Relationship between parameters

Four restrictions to the feasible set of values of the circuit parameters can be derived through the analysis performed in the previous section. The first one comes from the disappearance of the highest output level considering that $V_{in}=V_{in}^H$. We said that the critical situation occurs when both solutions around $V_{out} \cong V_{bias}^H - (V_{p1})_L$ collapse.

According to Fig. 5b it is mandatory that around this voltage, the first peak current of the load is above the current of the driver in order to ensure the existence of the solutions marked with red points. Thus, an expression concerning to a maximum *FF* is obtained,

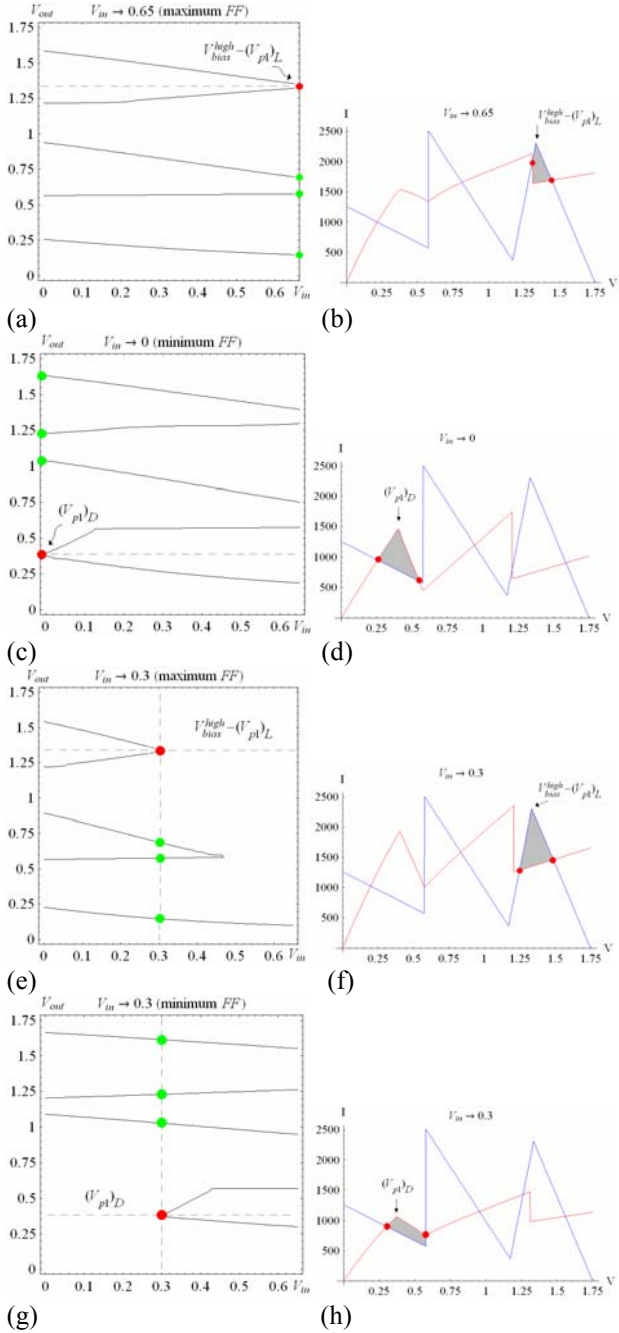

**Figure 3**. . $V_{in}$-$V_{out}$ plots pointing out the critical (in red) and the non-critical (in green) solutions, and load curve for (a), (b) $V_{in} = V_{in}^H$, (c), (d) $V_{in} = V_{in}^L$ and (e)-(h) $V_{in} = V_{in}^M$

$$g_D \left[ V_{bias}^H - (V_{p1})_L \right] + FF \cdot I_T \left[ V_{in}^H, V_{bias}^H - (V_{p1})_L \right] < f_B I_p \quad (2)$$

When $V_{in} = V_{in}^L$, the decision is taken around $V_{out} \cong (V_{p1})_D$ and must guarantee that the peak current of the driver is under the current of the load. The red points in Fig. 5d corresponds to the solutions obtained near this output voltage when the form factor of the transistor has been sized over a minimum value,

$$f_Z I_p + FF \cdot I_T \left[ V_{in}^L, (V_{p1})_D \right] > g_L \left[ V_{bias}^H - (V_{p1})_D \right] \quad (3)$$

For the medium input voltage, $V_{in} = V_{in}^M$, two conditions can be derived. A maximum value of *FF* is obtained by considering that the first peak current of the load must be above the current through the driver (Fig. 5d and 5e),

$$g_D \left[ V_{bias}^H - (V_{p1})_L \right] + FF \cdot I_T \left[ V_{bias}^H - (V_{p1})_L, V_{in}^M \right] < f_B I_p \quad (4)$$

Finally, the last relationship between parameters comes by forcing the first peak current of the driver to be greater than the current through the load,

$$f_Z I_p + FF \cdot I_T \left[ (V_{p1})_D, V_{in}^M \right] > g_L \left[ V_{bias}^H - (V_{p1})_D \right] \quad (5)$$

### 4.2 Results

Expressions (2-5) have been employed to analyze how the DC operation of a ternary inverter is modified when some key parameters are changed and thus, the range of feasible values of *FF* for a holding preserving in a ternary inverter can be studied. RTDs/HFET from the LOCOM technology, with $V_{bias}^H = 1.75$, $V_{in}^L = 0V$, $V_{in}^M = 0.3V$ and $V_{in}^H = 0.65V$ have been used. It can be proved that the constraints meaning a minimum value of *FF* provide negatives values (this limit will be taken in zero).

We have analyzed two different situations. In Figure 4a, feasible sets of values of *FF* for holding preserving inverters have been depicted. A constant difference ($\Delta$) between the RTD area factors, (that is, $f_Y = f_Z + \Delta$, $f_B = f_Z + 2\Delta$ and $f_A = f_Z + 3\Delta$ with $f_Z = \{0.2, 0.4, 0.6\}$ has been considered). Figure 4a shows the results which have been obtained being the regions indicated.

The second set of correct values of *FF* is obtained by considering $\Delta_1$ as the difference between $f_Z$ and $f_Y$ and $f_B$ and $f_A$, and $\Delta_2 = f_B - f_Y$, in Figures 4b and 4c ($f_Z = 0.6$). Fig. 4b and 4c depict the maximum value of *FF* against $\Delta_1$ and $\Delta_2$ respectively. Figure 4c has been used to check that both structures described in Fig. 2a and 2b ($\Delta_1 = 0.1$ and $\Delta_2 = 0.4$) operate as the HSPICE simulations performed indicate (black points).

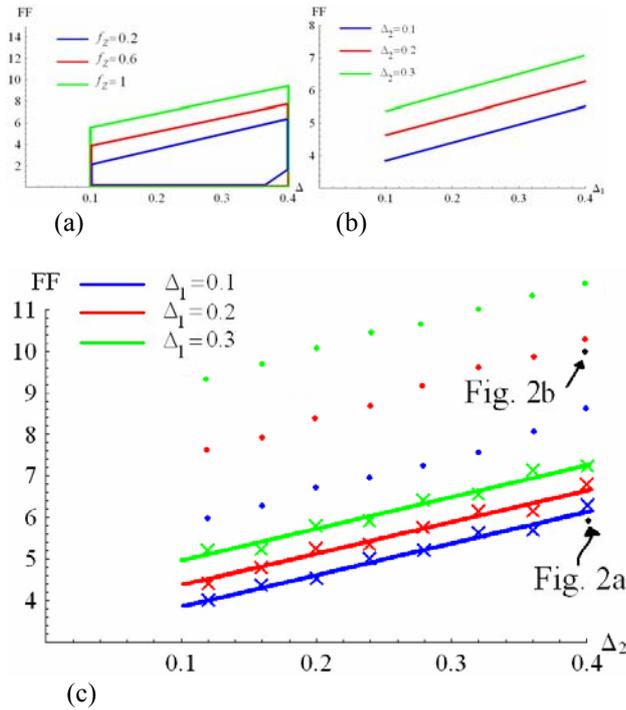

**Figure 4**. Feasible set of values of *FF* vs (a) $\Delta = f_Y - f_Z = f_B - f_Y = f_A - f_B$, (b) $\Delta_1 = f_A - f_B = f_Y - f_Z$ for $\Delta_2 = f_B - f_Y = \{0.1$ (blue), $0.2$ (red), $0.3$ (green)$\}$ and (c) $\Delta_2 = f_B - f_Y$ for $\Delta_1 = f_A - f_B = f_Y - f_Z = \{0.1$ (blue), $0.2$ (red), $0.3$ (green)$\}$, marking with points the critical conditions obtained from HSPICE and with crosses the conditions associated to the analysis taken individual current-voltage characteristic of the RTDs.

The red, blue and green crosses, depicts the maximum *FF* value when we consider HSPICE simulations using piecewise linear RTDs, which are really close to the constraints described in Eq. (2) to (5). Finally, the blue, red and green points depict the maximum *FF* obtained after performing HSPICE simulations with LOCOM RTDs for $\Delta_1=0.1$, $\Delta_1=0.2$ and $\Delta_1=0.3$ respectively. It is easy to check that the constraints obtained are more restrictive than the real situation, thus they can be used to ensure that the structure designed preserves the holding property.

## 5. CONCLUSIONS

Holding preserving is a key property that must verify sample and hold circuits, which can only be guaranteed by a proper choice of circuit parameters. A procedure to obtain the relation between those parameters in order to achieve a correct operation has been described by using piecewise linear current voltage characteristic of two series connected RTDs. HSPICE simulations and results after considering a linear approach of each RTD individually show a very good agreement with the results obtained.